International Journal of Accounting and Financial Reporting
ISSN 2162-3082
2018, Vol. 8, No. 4

# Parameters of Profitability: Evidence From Conventional and Islamic Banks of Bangladesh

K. M. Golam Muhiuddin

Vice-Chancellor, International Islamic University Chittagong

Kumira, Chittagong-4318, Bangladesh

Nusrat Jahan (Corresponding author)

Adjunct Faculty, Department of Business Administration

International Islamic University Chittagong

Kumira, Chittagong-4318, Bangladesh

Former Assistant Professor, CIU Business School

Chittagong Independent University

E-mail: mrs_ashfaque@yahoo.com

**Abstract**

This paper evaluates the commercial banks of Bangladesh in terms of profitability dimension of performance and also examines the impact of selected determinants and banking system on this dimension of performance. Evaluation of trend in profitability of listed commercial banks of Bangladesh reveals that, on an average, profitability is exhibiting a decreasing trend over the selected period; however, the profitability performance of Islamic banks remained rather high compared to Conventional banks. Profitability measured by Return on Asset is found to be significantly affected by the bank-specific factors, industry-specific factor and the banking system. However, macro-economic factors evidently have no significant impact on profitability of commercial banks of Bangladesh.

International Journal of Accounting and Financial Reporting
ISSN 2162-3082
2018, Vol. 8, No. 4

## 1. Introduction

Banks are financial institution that acts as financial intermediaries to pool financial resources from surplus units and allocate them to deficit units for investment purposes which ultimately results in economic growth of a country. Rapid financial deregulation, consolidation, technological advancements and financial innovation are forces that have made revolutionary changes in the banking sector worldwide in past few decades. Hence, banks are undertaking new approaches to develop better financial solutions for their customers taking into account the forces in effect. Financial innovation leads to development of new financial product or instrument or an entirely new financial intermediary system. Technological changes and competition pushes financial institution to change their system of intermediation for better efficiency into one that matches the ongoing changes in the banking sector. Banking industry worldwide has given rise to a number of different banking systems while performing the same intermediary function. The banking sector of Bangladesh comprises four categories of scheduled banks. These are state-owned commercial banks (SCBs), state-owned development financial institutions (DFIs), private commercial banks (PCBs) and foreign commercial banks (FCBs). Alongside the Conventional interest based banking system, Bangladesh entered into an Islamic banking system in 1983. In 2015 out of the 56 commercial banks in Bangladesh, 8 operated as full-fledged Islamic banks. The Islamic banking industry continued to show strong growth since its inception in tandem with the growth in the economy. This is reflected by the increased market share of the Islamic banking industry in terms of its assets, financing and deposits in the total banking system. Individuals and enterprises in Bangladesh largely depend on banking institutions for the purpose of financial intermediation. Most financial market participants, such as - borrowers, lenders, investors, bank shareholders or regulators, are familiar with and engage or transact in commercial banks; whereby banks perform the financial intermediation process by charging interest on advances extended to borrowers and paying interest to compensate depositors for providing the source of funding with the objective of making a profit on the spread or margin in between. In Bangladesh where majority of the population is Muslim, people are fast turning to Islamic banking as an alternative mode of banking. Islamic banking is a form of modern banking in which the financial intermediation process is based on a number of Islamic, legal and economic principles developed in the first centuries of Islam that uses risk-sharing methods instead of interest-based financing (Van Schaik, 2001).

Therefore, in the presence of a bank dominated financial system where the dependence on bank for financial intermediation is substantial, the banking sector can be considered as fertile ground for research; since for growth and development of economy of Bangladesh banking sector serves as an important catalyst. The debt market of Bangladesh is still undeveloped hence commercial banks still serve as a main provider of funds to individuals and enterprises. Therefore, stability of the banking sector of Bangladesh is of paramount importance to the smooth functioning of financial system of this country. Economies that have a profitable banking sector are better able to withstand negative shocks and contribute to the stability of the financial system (Athanasoglou et al, 2006). Besides, Maximization of profit is a short-run goal of a firm and firms in reality are more eager to meet its short-term earnings target than long-run goal of shareholders wealth maximization. However, successfully meeting the short-run goals will eventually be leading to meeting the long-run financial goal of a firm. Hence, this notion indicates that it is important to evaluate the profitability of banking sector and also to understand the factors which drive the profitability of this sector in Bangladesh.

## 2. Review of Literature on Bank Profitability and Its Determinants

Profitability is an important measure of financial performance for any bank. Analysis of profitability of a bank provides an insight into the effective utilization of its assets and both debt and equity fund. Profitability is considered as an index of operational efficiency of banks (Shrivastava, 1979). There are a lot of literatures that have evaluated performance of banks in terms of profitability estimation by using financial ratios which are basically return on Asset (ROA), Return on Equity (ROE) and Net Interest Margin (NIM). The central bank of Bangladesh, i.e. Bangladesh Bank, also evaluates profitability in terms of these three measures. Return on Asset (ROA) which is a ratio between earnings after tax and total assets reflects the ability of management to generate income on a given amount of total assets. According to DuPont analysis, ROA indicates both income management and cost management of banks by including both asset utilization ratio and net profit margin ratio. Asset utilization ratio is measured by total operating revenue to total assets and net profit margin ratio is measured by net earnings to total operating revenue. Therefore, ROA assesses how efficiently bank is managing its revenues and expenses; and also reflects the bank managements' ability to generate profits by using the available financial and real assets (Clark et al. 2007). Evaluating the performance of banks through profitability indicators and investigating the determinants of banks profitability has been one of the popular topics among researchers in banking studies. Till date, researchers have managed to examine and identify various factors that have a significant influence on banks profitability. The literature divides the determinants of bank profitability into two categories, namely internal and external. Internal determinants of profitability which are within the control of bank management are basically financial statement variables. External variables are those factors that are considered to be beyond the control of management of a bank but reflect the economic and legal environment that influences the operation and performance of financial institutions. A number of explanatory variables have been proposed by researchers from both categories, according to the nature and purpose of each study. The researches undertaken on determinants of profitability either focused on cross-country or individual countries banking system. Past studies examined either internal or combinations of internal and external determinants of bank profitability. The empirical results of these studies vary significantly; since variables, datasets and environments differ according to scope and aim of each study. Moreover, based on some common criterion, past researches have also allowed a further categorization of the determinants. Studies have shown that several internal factors such as overheads, liquidity ratio, leverage ratio, capital ratio, cost to income ratio, efficiency ratio, credit risk, solvency risk, operating expense, deposits, and bank size are considered as bank-specific factors. On the other hand, external factors such as GDP, inflation, interest rate and the exchange rate are considered as macro-economic factors; and market concentration and market share as industry-specific factors. The following discussion lists few researches that were aimed at evaluating the performance of commercial banks of Bangladesh.

Hassan (1999) examined and compared the performance of Islamic Bank Bangladesh Limited with other private banks in Bangladesh between 1993 and 1994. Though the duration of this study was short, however, the result revealed that in terms of deposits growth and investments growth, performance of Islamic Bank Bangladesh Limited was better than that of selected private commercial banks. Apart from that, the researcher found that the key Islamic financial products, Mudharabah and Musyarakah were not well developed to compete with the products of private commercial banks. Siddique and Islam (2001) undertook a study on commercial banks of Bangladesh over the financial year of 1980 to1995. This study revealed that in every aspect, transnational banks had a commendable performance in Bangladesh. But comparison among other groups of banks which are Nationalized Commercial Banks (NCBs), Specialized Banks (SPBs) and Private Commercial Banks (PCBs); PCBs had favorable achievement over others in terms of profit. On the other hand, Specialized Banks in Bangladesh had a very poor performance and their meager activity affected the overall banking sector's performance. Chowdhury (2002) in his study emphasized that performance of banks requires knowledge about the profitability and its relationships with variables like market size and risk. This study concluded that the banking industry in Bangladesh is experiencing a major transition over the last two decades. The author recommended that the banks that can endure the pressure arising from both internal and external factors prove to be profitable. Hasan and Omar (2006) made a comparative performance analysis between state-owned and privately-owned commercial banks of Bangladesh over the period of 2006 to 2010. Return on Asset and Return on Equity (ROE) were used to measure profitability; and Net profit and net asset relative to total employment and total number of branches were used to measure operating efficiency. The results suggested that state-owned banks are as efficient as private banks; but private banks reported to have much higher mean values for the selected measures.

Jahangir et al. (2007) stated that the traditional measure of profitability through return on stockholder's equity is quite different in banking industry from any other sector of business, whereas loan-to-deposit ratio works as a very good indicator of banks' profitability as it depicts the status of asset-liability management of banks. Besides, banks' market size and market concentration index along with return on equity and loan-to-deposit ratio also grab the attention while analyzing the banks' profitability. Chowdhury and Islam (2007) stated that deposit, and loans and advances of nationalized commercial banks (NCBS) are less sensitive to interest rate changes than those of specialized commercial banks (SCBs). They also suggested that higher return on equity (ROE) is also noticeable as it is the primary indicator of bank's profitability and financial efficiency. Nimalathasan (2008) undertook a comparative study of financial performance of banking sector in Bangladesh using CAMELS rating system. This study was done on 6562 Branches of 48 Banks in Bangladesh over the financial year of 1999 to 2006. This study revealed that out of 48 banks, 3 banks were rated 01 or Strong, 31 banks were rated 02 or satisfactory, 7 banks were rated 03 or Fair, 5 banks were rated 04 or Marginal and 2 banks obtained 05 or unsatisfactorily rating. 1 Nationalized

Commercial Bank (NCB) had unsatisfactorily rating and other 3 NCBs had marginal rating.

Chowdhury and Ahmed (2009) investigated the performance of private commercial banks and revealed that all the commercial banks are able to achieve a stable growth of branches, employees, deposits, loans and advances, net income, earnings per share during the period of 2002 to 2006. Rushdi (2009) compared the performance of Islamic Bank Bangladesh Limited with Janata bank Limited in terms of accounting profitability, partial productivity and total factor productivity over the period of 1983 to 2006. This study confirms that the IBBL performed excellently in terms of labor and capital productivity and TFP over the studied period. Sufian and Habibullah (2009) reported in their study that bank specific characteristics, in particular, loan intensity, credit risk and cost have positive and significant impacts on profitability of Bangladeshi banks, while non-interest income exhibits negative relationship with bank profitability. This study found that size has a negative impact on return on average equity (ROAE) while it has positive effect on return on average assets (ROAA) and net interest margin (NIM). Safiullah (2010) emphasized on the financial performance analysis of Conventional and Islamic banks to measure their superiority. The research result indicated that interest-based conventional banks are doing better performance than interest-free Islamic banks in terms of commitment to economy and community, productivity and efficiency. However, performance of interest-free Islamic banks in business development, profitability, liquidity and solvency is superior to that of interest-based conventional banks.

Sarker and Saha (2011) investigated the performance of NCBs, PCBs, FCBs and SCBs through highlighting their profitability, branch productivity, employee productivity and overall productivity and also by using SWOT mix during the period of 2000 to 2009. Sufian and Kamarudin (2012) identified bank-specific characteristics and macro-economic determinants of profitability of 31 commercial banks over the period of 2000 to 2010. This study suggested five bank-specific determinants that are important in influencing profitability are capitalization, non-traditional activities, liquidity, management quality, and size of the bank. Besides, this study found that three macro-economic determinants significantly influence profitability including growth in GDP, inflation and market concentration. Jahan (2012) evaluated randomly selected six commercial banks of Bangladesh by using widely used indicators of banks' profitability, which are Return on Asset (ROA), Return on Equity (ROE) and Return on Deposit (ROD). This study investigated the impact of efficiency ratio, asset utilization ratio, asset size and ROD as a determinant of banks' profitability measured by ROA. The result of regression analysis found that operational efficiency, asset size and ROD to be positively related and asset utilization to be negatively related to ROA, but these associations were statistically insignificant.

Haque (2013) investigated the financial performance of five private commercial banks in Bangladesh for the period of 2006 to 2011 under four dimensions: (1) profitability (2) liquidity (3) credit risk and (4) efficiency. The study concluded that there is no specific relationship between the generation of banks and its performance. The performance of banks are dependent more on the management's ability to formulate strategic plans and the efficient implementation of such strategies. Paul et al. (2013) evaluated the profitability and liquidity of Conventional and Islamic banks of Bangladesh for the period of 2008 to 2012. They found that profitability performance of Islamic banks' is better than Conventional banks; however, there is no significant difference in liquidity between these two types of banks. Jahan (2014) evaluated six Islamic banks of Bangladesh in terms of profitability indicators ROA, ROE and ROD. This study also evaluated the relationship of ROA with asset utilization, operational efficiency and Return on Deposit (ROD). The result revealed that asset utilization and operational efficiency to be significantly related to ROA but failed to establish any significant relationship with ROD. Rahman et al. (2015) investigated the determinants of profitability of 25 commercial banks in Bangladesh for the period of 2006 to 2013. Three different measures of profitability namely return on assets (ROA), net interest margin over total assets (NIM) and return on equity (ROE) are used in this study. The findings suggested that capital strength and loan intensity have significant positive impact on profitability while cost efficiency and off-balance sheet activities have significant negative impact on banks profitability. Non-interest income, credit risk and growth of GDP are found to be important determinants of NIM.

The above review of past literatures indicates that several researches exist with respect to profitability evaluation of commercial banks. However, the number of researches that are all encompassing i.e. examines the impact of bank-specific, industry-specific and macro-economic factors on profitability is very limited in the context of Bangladesh. Besides, comprehensive literature on comparative performance analysis of Conventional and Islamic banks of Bangladesh is seemingly scarce. Furthermore, the research examining the impact of banking system on profitability of commercial banks of Bangladesh is also inadequate.

## 3. Objectives of the Study

The main objectives of this study are as follows:

First, to evaluate the trend of commercial banks' profitability over the selected period of study.

Second, to identify which banking system is boasting superiority in terms of profitability performance.

Third, to explore what bank-specific, industry-specific and macro-economic factors are affecting the profitability performance.

Fourth, to examine what impact the different banking system exert on the profitability performance.

## 4. Research Methodology

### *4.1 Philosophical Position of This Study*

The research design for this study is empirical and the researcher's view is positivist and post-positivist where the aim is explanation, prediction and control; analyzed through quantitative methods. Quantitative data are collected through structured procedure from a large sample; and statistical tools and techniques used also reflect the philosophy of positivism. The present study follows deductive approach to carry out the research;

hypotheses are developed and tested to explain causal relationship and make generalizations. Both descriptive and inferential statistics are used and parametric tests are applied for data analysis. Descriptive statistics measures used includes arithmetic mean, minimum, maximum, standard deviation and trend analysis. Parametric inferential statistics applied is pooled Ordinary Least Square (OLS) regression with Panel Corrected Standard Error (PCSE) estimation.

*4.2 Population, Sample and Time Period of Study*

There are 56 commercial banks in Bangladesh of which 48 are Conventional and 8 are Islamic banks. Among these 56 commercial banks, 29 are listed in stock exchanges of Bangladesh. The sample comprises of these 29 listed commercial banks, of which 23 are Conventional and 6 are Islamic banks and the data collected covers a time period between 2011 and 2015.

*4.3 Sources and Type of Data*

The data are obtained from individual commercial bank's audited annual reports, annual report of Bangladesh Banks and also the financial stability report published by Bangladesh Bank. The nature of this dataset is pooled time-series cross-sectional (TSCS) data. Time-series cross-sectional (TSCS) data resemble panel data but typically have the opposite structure of panel data. Pooled data are characterized by having repeated observations (most frequently years) on fixed units (i.e. bank). This means that pooled arrays of data are one that combined cross-sectional data on 'n' spatial units (i.e. 29 banks) and 't' time periods (i.e. 5 year time periods) to produce a data set of 'n $\times$ t' (i.e. 29 X 5 = 145) observations for this study (Podesta, 2000; Beck, 2006).

*4.4 Measurement of Determinants*

The internal and external determinants of performance of commercial banks have been identified through the review of past studies on bank performance. The bank-specific determinants are micro or internal factors that originate from the balance sheet and profit and loss account; and these factors are influenced by the management decisions and policy objectives (Staikouras and Wood, 2011). The bank-specific determinants of profitability considered in this study are size to account for the economies of scale and proxy for size is log of total assets; productivity to account for efficiency and technological change and proxy used is Malmquist Productivity Index of Total Factor Productivity estimated by using two inputs - interest expense and non-interest expense, and two outputs- interest income and non-interest income; credit risk to account for loans in default and proxy for credit risk is Non-Performing loan (NPL) ratio; liquidity risk to account for insolvency and measured by total loan to total deposit ratio; operating inefficiency to account for cost management and proxy is operating expenses to net interest income ratio; and finally capital structure to account for financial risks and solvency and measured by total equity to total asset ratio.

The external determinants are comprised of macro-economic and industry-specific variables, which are outside of the prerogative of bank-specific decisions and policies (Athanasoglou et al., 2008; Staikouras and Wood, 2011). The market or industry specific variable that is taken into consideration is market concentration since it is expected to have significant influence on performance of commercial banks. Concentration is linked to the degree of competition and highly concentrated banking industry is considered to have lower degree of competition; and therefore conducive for implicit or explicit collusion among banks to earn higher than normal profits (Bain, 1951). Herfindahl–Hirschman Index (HHI) for loans is used in this study for measuring concentration of banking industry. Bank performance is also likely to be influenced by the environment in which they operate. Hence, to account for the relationship between performance of banks and macro-economic state of the economy, macro-economic factors that are considered in this study are inflation and GDP growth rate. To avoid the non-stationarity characteristics of time series macro-economic variable, natural logarithm of real inflation rate is used as a proxy for inflation rate. This study uses real GDP growth rate to capture upswings and downswings of economy over the selected period.

*4.5 Measure of Profitability: Return on Assets (ROA)*

The return on asset (ROA) is a substantial performance measure because it is directly related to the profitability of banks (Kosmidou, 2008; Sufian and Habibullah, 2009). ROA have been used in most of the studies for the measurement of profitability of the banks. ROA is a ratio calculated by dividing the net income over total assets. ROA measures the profit earned per dollar of assets and reflects how well bank management uses the bank's real investments resources to generate profits (Naceur, 2003; Alkassim, 2005).

*4.6 Statistical Tools and Programs Used for Evaluating Profitability Performance*

This study uses descriptive statistical measures such as mean, minimum, maximum and standard deviation to summarize the dataset included in evaluation of profitability performance. This study applies trend analysis for evaluating the profitability performance of commercial banks over the period of 2011 to 2015. The trend analysis of ROA for all listed commercial banks with a breakdown for two banking group is presented through the tables and charts for evaluation and comparison. Furthermore, yearly average and five years average ROA , standard deviation of five years average ROA and coefficient of variation of yearly average ROA are used to evaluate the profitability performance of all listed commercial banks and also the Conventional and Islamic banking group. The maximum and minimum values of ROA of all listed banks and standard deviation from yearly average ROA and coefficient of variation of yearly ROA of all banks are also used to assess the profitability performance of commercial banks. The data collected for performance evaluation of commercial banks are tabulated and charts are created using Excel. Description of dataset and regression analysis is carried out using statistical software 'Stata'.

*4.7 Hypothesis Formulation for Evaluating Profitability Performance*

To show the impact of external and internal determinants and banking system on profitability of commercial banks of Bangladesh, two null hypotheses are formulated, which are to be tested through the current study, are explained below.

H1o: There is no impact of explanatory variables on profitability of commercial banks of Bangladesh.

H2o: There is no impact of banking system on profitability of commercial banks of Bangladesh.

Second null hypothesis is an extension of first null hypothesis, which accommodates examination of bank type to show the impact of Conventional and Islamic banks on profitability performance. The first null hypothesis evaluates the affect of selected firm-specific, industry-specific and macro-economic determinants on profitability performance of commercial banks of Bangladesh. The first null hypothesis is tested by examining the significance of linear regression model's beta coefficient for each determinant. If the calculated probabilities of all beta coefficients of selected determinants are less than 0.05 level of significance, then the first null hypothesis (H1o) will be rejected. The second null hypothesis is tested by examining the significance of beta coefficient of dummy variable indicating the type of bank. If the calculated probability of beta coefficient for dummy variable is less than 0.05 level of significance, then the second null hypothesis (H2o) will be rejected.

*4.8 Regression Model Specification for Profitability Performance*

Pooled Ordinary Least Square (OLS) regression estimation exclusively cannot take account of probable contemporaneous correlation of the errors and heteroskedasticity of time-series cross sectional (TSCS) data used for the current study. Besides fixed effect estimator fails to identify the effect of time-invariant variables like banking system on the regression line. In fixed effect estimation, the parameter reported as 'constant' indicates the average fixed effect not the individual fixed effect particular to each type of bank. However, using Dummy Variable in Ordinary Least Square (OLS) regression estimator provides a good way to understand fixed effect of two different types of bank on profitability performance. So, following the study of Mahmud et al. (2016), this study opt for using a fixed effect estimation by pooled OLS with computation of a panel-corrected covariance matrix of the coefficient estimates that automatically corrects heteroskedasticity and autocorrelation problem of a linear model. A set of N-1 dummy variables are added in OLS with panel corrected covariance matrix estimation to identify the entity effect that means the dummy variables absorbs the effect particular to each type of bank. Therefore, the following pooled ordinary least square (OLS) estimator with fixed entity effect and panel corrected standard error estimation (PCSE) is used to show the impact of selected determinants and two different banking systems on the profitability of all listed commercial banks:

$Y_{it} = \beta_0 + \beta_1 size_{it} + \beta_2 crisk_{it} + \beta_3 liq_{it} + \beta_4 opeff_{it} + \beta_5 capstr_{it} + \beta_6 TFP_{it} + \beta_7 hhi_{it} + \beta_8 infl_{it} + \beta_9 GDP_{it} + \beta_{10} Dconv_{it} + \beta_{11} Dislm_{it} + \mu_{it}$

Where,

$Y_{it}$ = ROA of i-th bank in time t

$size_{it}$ = size of i-th bank in time t

$crisk_{it}$ = credit risk of i-th bank in time t

$liq_{it}$ = liquidity risk of i-th bank in time t

$opeff_{it}$ = operating inefficiency of i-th bank in time

$capstr_{it}$ = capital structure of i-th bank in time t

$TFP_{it}$ = total factor productivity of i-th bank in time t

$hhi_{it}$ = market concentration of i-th bank in time t

$infl_{it}$ = inflation rate sustained by i-th bank in time t

$GDP_{it}$ = GDP growth rate sustained by i-th bank in time t

$Dconv_{it}$ =1 if i-th bank in time t is Conventional bank otherwise 0

$Dislm_{it}$ = 1 if i-th bank in time t is Islamic bank otherwise 0

$\mu_{it}$ = the normal error term

## 5. Findings and Analysis

### 5.1 Descriptive Statistics for Profitability Performance

Table 1 details the overall mean, standard deviation, maximum and minimum values of dependent and explanatory variables used in evaluation of profitability performance of commercial banks of Bangladesh. The table also presents the breakdown of variables into a between and within variations. Table 1 reports that total number of observation N=145, number of banks n=29 and time period t=5 (i.e. 2011-2015). The overall is calculated for 145 observations, between values are calculated across 29 banks and within values are calculated across five years for each bank. In Table 1, 'bn' indicates bank's name coded with numerical values 1 to 29 and 'bc' indicates category of banks, also coded with numerical values; 1 for Conventional and 2 for Islamic banks and year indicates the selected period of study.

Table 1. Descriptive statistics for profitability performance

| **Variable** | | **Mean** | **Std.Dev.** | **Min.** | **Max** | **Observations** |
|---|---|---|---|---|---|---|
| **roa** | Overall | 0.122132 | 0.0074489 | 0.0008 | 0.07 | N =145 |
| | Between | | 0.0045476 | 0.0038 | 0.02444 | n = 29 |
| | within | | 0.0059481 | -0.0026268 | 0.0577732 | T = 5 |
| **TFP** | Overall | 0.9806897 | 0.1953504 | 0.323 | 2.993 | N =145 |
| | Between | | 0.0541764 | 0.9044 | 1.2282 | n = 29 |
| | within | | 0.1879048 | 0.0754896 | 2.74549 | T = 5 |
| **size** | Overall | 10.97538 | 1.121342 | 5.03 | 12.19 | N =145 |
| | Between | | 1.130193 | 5.224 | 11.82 | n = 29 |
| | within | | 0.1247141 | 10.59738 | 11.85738 | T = 5 |
| **crisk** | Overall | 0.0567062 | 0.0477855 | 0.0095 | 0.3 | N =145 |
| | Between | | 0.0456773 | 0.022 | 0.246 | n = 29 |
| | within | | 0.0159796 | 0.0010262 | 0.1402262 | T = 5 |
| **liq** | Overall | 0.8153138 | 0.0777499 | 0.5615 | 1.029 | N =145 |
| | Between | | 0.0646639 | 0.62128 | 0.92014 | n = 29 |
| | within | | 0.0444947 | 0.6885138 | 0.9620538 | T = 5 |
| **opeff** | Overall | 1.159668 | 1.113493 | -3.27 | 6.14 | N =145 |
| | Between | | 0.9204991 | -0.098 | 5.064 | n = 29 |
| | within | | 0.6450461 | -2.012332 | 4.887668 | T = 5 |
| **capstr** | Overall | 0.0850483 | 0.0201592 | 0.037 | 0.154 | N =145 |
| | Between | | 0.0188407 | 0.043 | 0.1334 | n = 29 |
| | within | | 0.0078284 | 0.0646483 | 0.1156483 | T = 5 |
| **hhi** | Overall | 1438.84 | 174.728 | 1269 | 1670 | N =145 |
| | Between | | 0 | 1438.84 | 1438.84 | n = 29 |
| | within | | 174.728 | 1269 | 1670 | T = 5 |
| **infl** | Overall | -1.120269 | 0.0541727 | -1.187087 | -1.055517 | N =145 |
| | Between | | 0 | -1.120269 | -1.120269 | n = 29 |
| | within | | 0.0541727 | -1.187087 | -1.055517 | T = 5 |
| **GDP** | Overall | 0.0632 | 0.0022348 | 0.06 | 0.065 | N =145 |
| | Between | | | 0.0632 | 0.0632 | n = 29 |
| | within | | | 0.06 | 0.065 | T = 5 |
| **bn** | Overall | 15 | 8.395601 | 1 | 29 | N =145 |
| | Between | | 8.514693 | 1 | 29 | n = 29 |
| | within | | 0 | 15 | 15 | T = 5 |
| **bc** | Overall | 1.206897 | 0.4064848 | 1 | 2 | N =145 |
| | Between | | 0.4122508 | 1 | 2 | n = 29 |
| **year** | within | | 0 | 1.206897 | 1.206897 | T = 5 |
| | Overall | 2013 | 1.419119 | 2011 | 2015 | N =145 |
| | Between | | 0 | 2013 | 2013 | n = 29 |
| | within | | 1.419119 | 2011 | 2015 | T = 5 |

*5.2 Trend in Profitability (ROA) of Commercial Banks*

This study evaluates the profitability performance of listed commercial banks inclusive of Conventional and Islamic banks by the profitability ratio ROA. The evaluation of bank's profitability over the study period is done by trend analysis of ROA and comparison of ROA among two banking groups and all listed commercial banks. The following Table 2 presents the annual average ROA for all listed banks and each of two different banking groups - Conventional banks and Islamic banks.

Table 2. Trend and comparison of ROA

| Year | Banking Sector Yearly Average ROA | Conventional Banks Yearly Average ROA | Islamic Banks Yearly Average ROA |
|---|---|---|---|
| **2011** | 0.015 | 0.017 | 0.0179 |
| **2012** | 0.006 | 0.0117 | 0.0148 |
| **2013** | 0.009 | 0.0099 | 0.0109 |
| **2014** | 0.006 | 0.0105 | 0.0104 |
| **2015** | 0.008 | 0.0099 | 0.0097 |

The results presented in Table 2 and Figure 1 reports that yearly average ROA across the reported five year period for Islamic banks and Conventional banks remains higher than overall banking sector average. However, the annual average ROA of Islamic banks remains higher than Conventional banks till 2013 but declined a little during 2014 and 2015. In Figure 1, it is apparent that though both conventional and Islamic banks have ROA above the banking sector average but their profitability is showing a declining trend over the study period.

International Journal of Accounting and Financial Reporting
ISSN 2162-3082
2018, Vol. 8, No. 4

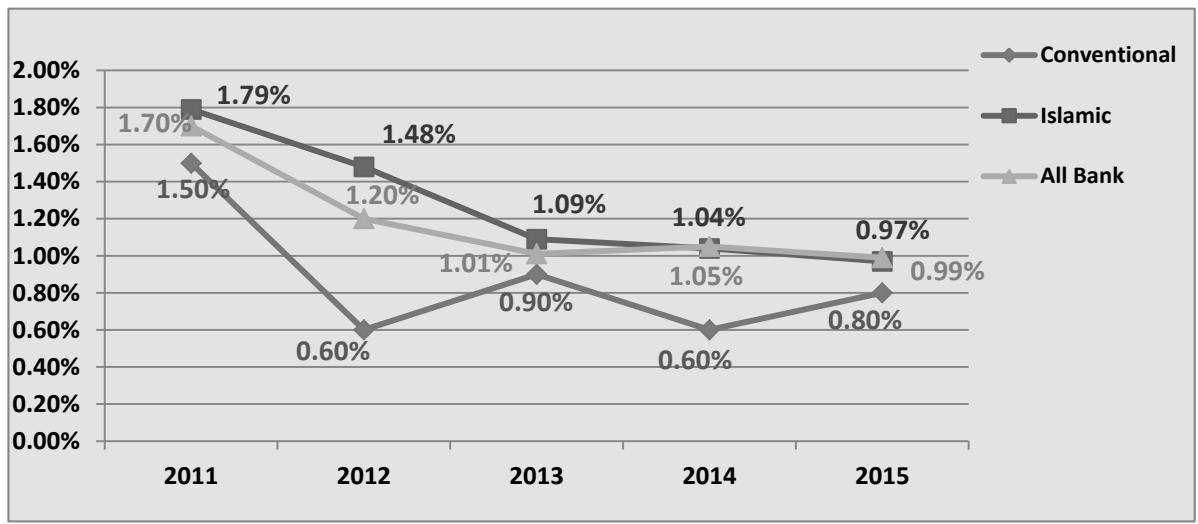

Figure 1. Trend and comparison of ROA

Table 3 reports five years average ROA for both banking groups and all listed commercial banks and also presents variation of year to year mean ROA by standard deviation and coefficient of variation. The maximum and minimum values of yearly average ROA of all listed banks and their standard deviation and coefficient of variation are also reported in this table. The result shows that Islamic banks five years average ROA, which is 1.3%, is higher than the total listed banks' five years average ROA of 1.2% However, the Conventional banks and all listed banks' five years average ROA is identical that is 1.2%.

Table 3. Summary of annual ROA mean

| Banks | 2011 | 2012 | 2013 | 2014 | 2015 | Five Year's Average | SD | CV |
|---|---|---|---|---|---|---|---|---|
| **Conv.** | 0.017 | 0.0117 | 0.0099 | 0.0105 | 0.0099 | 0.012 | 0.003 | 0.2541 |
| **Islamic** | 0.0179 | 0.0148 | 0.0109 | 0.0104 | 0.0097 | 0.013 | 0.004 | 0.2747 |
| **All Banks** | | | | | | | | |
| **Max** | 0.0401 | 0.07 | 0.0168 | 0.0236 | 0.0208 | | | |
| **Min** | 0.0061 | 0.001 | 0.0019 | 0.001 | 0.0008 | | | |
| **Yearly Average** | 0.017 | 0.012 | 0.0101 | 0.0105 | 0.0099 | 0.012 | 0.003 | 0.2548 |
| **SD** | 0.007 | 0.012 | 0.0038 | 0.0048 | 0.0045 | | | |
| **CV** | 0.408 | 0.996 | 0.379 | 0.4533 | 0.4537 | | | |

Furthermore, the variation from five years average ROA reported by standard deviation which is 0.3% is identical for both Conventional banks and all listed commercial banks. Moreover, the reported year to year variation in ROA of Conventional banks reported by coefficient of variation of 25.4% is also similar to that of all listed commercial banks. Though Islamic banks five years average ROA, which is 1.3%, is higher than Conventional banks and all listed commercial banks ROA of 1.2% but the variation from five years average ROA as reported by standard deviation of ROA is higher for Islamic banks compared to Conventional banks and all listed commercial banks. In addition, the coefficient of variation of 27.5% indicates that year to year variation of ROA of Islamic banks is also high compared to Conventional banks and all listed banks. Therefore, profitability of Islamic banks is more volatile compared to Conventional banks. Overall, the trend analysis of ROA and comparison of ROA among banking groups and all listed commercial banks shows that both banking groups reported to have above average profitability over the study period.

*5.3 Impact of Selected Determinants and Banking System on Profitability*

The result of regression is reported in Table 4 indicating that estimation is carried out for 145 observations on bank-year. Regression result indicates that, size, credit risk, operating efficiency, capital structure, HHI and liquidity risk are found to be statistically significant at 5% level of significance since "P" value are less than 0.05. Hence, this study rejects the first null hypothesis (H1o) and concludes that there exists significant effect of explanatory variables on profitability performance of commercial banks of Bangladesh.

Table 4. Result of regression estimation for profitability performance

| roa | Coefficient | Std. Error | Z | P > [Z] | [95% Conf. Interval] | |
|---|---|---|---|---|---|---|
| **size** | 0.0004756 | 0.0001324 | 3.59 | 0.000 | 0.0002161 | 0.000735 |
| **crisk** | - 0.0255517 | 0.0048779 | -5.24 | 0.000 | -0.0351121 | -0.0159913 |
| **opeff** | -0.0005283 | 0.0002042 | -2.59 | 0.010 | -0.0009286 | -0.000128 |
| **capstr** | 0.1357062 | 0.0146703 | 9.25 | 0.000 | 0.1069529 | 0.1644594 |
| **hhi** | 0.0000224 | 0.000011 | 2.04 | 0.042 | 8.52e-07 | 0.000044 |
| **infl** | -0.0287161 | 0.0293315 | -0.98 | 0.328 | -0.0862047 | 0.0287725 |
| **GDP** | -0.4213027 | 0.2927992 | -1.44 | 0.150 | -0.9951786 | 0.1525732 |
| **TFP** | 0.0023103 | 0.0017429 | 1.33 | 0.185 | -0.011058 | 0.0057264 |
| **Liq** | 0.0102031 | 0.0041197 | 2.48 | 0.013 | 0.0021286 | 0.0182777 |
| **2.bc** | 0.0016984 | 0.0005412 | 3.14 | 0.002 | 0.0006377 | 0.0027591 |
| **cons** | -0.0512469 | 0.034182 | -1.50 | 0.134 | -0.1182388 | 0.015745 |

Number of observations = 145

R-squared = 0.2890

Wald chi$^2$ = 328.89

Probability > chi$^2$ = 0.0000

Furthermore, the reported dummy variable for Islamic bank (2.bc) is also found to be statistically significant at 5% level of significance. Hence, this study also rejects the second null hypothesis (H2o) and concludes that there exists significant impact of banking system on profitability performance of commercial banks of Bangladesh. The impact of explanatory variables which are size, credit risk, operating efficiency, capital structure, HHI and liquidity risk on profitability, known as the slope effect, is identical for both Conventional banks and Islamic banks. But the dummy variable (2.bc) for Islamic bank which is an intercept dummy indicates that ROA of Conventional and Islamic bank is different. Therefore, the difference between profitability performance of Conventional and Islamic banks measured by ROA lies in the intercept dummy, but the explanatory variables i.e. the determinants remains same for both banking system. ROA of Islamic banks is on an average 0.0016984 unit more than the Conventional banks of Bangladesh. For goodness of fit of the model, $R^2$ that is, coefficient of determination is calculated, where $R^2$ being 0.2890, that means 28.90% of the variation in ROA is explained by the regression line. Wald Chi$^2$ statistics is used to test the significance of the model for overall goodness-of-fit. The calculated probability of Chi$^2$ test is less than 0.05, hence this regression model as a whole is found to be statistically significant at 5% level of significance. This indicates that the set of explanatory variables included in this regression model provides better fit to the current model than the intercept only regression model. Therefore, the result indicates there exist significant impact of explanatory variables and banking system on profitability performance of commercial banks of Bangladesh.

**6. Discussion on Profitability Performance of Commercial Banks of Bangladesh**

*6.1 Trend Analysis*

The profitability performance of all listed commercial banks of Bangladesh as indicated by yearly average ROA showed inconsistent trend in profitability over the study period of 2011 to 2015. The profitability performance all listed commercial banks remained highly volatile with the highest volatility in ROA reported by CV during the year 2012. The listed Conventional banks showed a trend in profitability similar to that of overall banking industry. Though, the listed Islamic banks reported to have a decreasing trend in profitability but their five years average was found to be a little higher compared to that of Conventional banks.

However, the volatility in profitability performance was reportedly higher for Islamic banks compared to Conventional banks. Therefore, these findings indicate that the profitability performance of Conventional banks was rather sound compared to inconsistent performance of Islamic bank.

*6.2 Impact of Bank-Specific Factors on Profitability (ROA)*

This study reports that firm-specific factors have significant impact on profitability performance of all listed commercial banks of Bangladesh. Firm-specific factors including size, capital structure and liquidity risk have positive association with ROA. Therefore, a unit increases in size indicative of increased scale of operation results in increment in ROA of commercial banks by the amount of beta coefficient which is 0.0004756 units. A unit increase in capital structure ratio signifying more usage of equity financing and less risk of insolvency, results in increment in ROA of commercial banks by 0.1357062 units. Furthermore, a unit increment in liquidity risk ratio representative of higher liquidity thus signifying less insolvency risk results in increase in ROA by 0.0102031 units. However, the other two significant firm-specific factors which are credit risk and operating inefficiency found to have inverse association with ROA which are theoretically consistent. The ratio used to calculate credit risk indicates that as the percentage of non-performing loan to total loan increase, leading to increment in credit risk affects earnings negatively. While an increase in the percentage operating cost compared to net interest income indicates higher value for operating inefficiency ratio, also signifying adverse effect on profitability. Hence, the inverse association of these two firm-specific factors with profitability of commercial banks of Bangladesh is theoretically appropriate. Thus a unit increase in credit risk reduces the ROA of commercial banks by 0.0255517 units, whereas a unit increase operating inefficiency ratio indicative of inefficient cost management results in lowering of the ROA by 0.0005283 units. The sign of beta coefficient indicate that impact of productivity (TFP) on profitability is positive as anticipated, thus suggesting profitability increases with the increase in productivity. Though the impact of TFP is statistically insignificant, however, the regression model for ROA as a whole is significant, thus all the determinants including TFP are providing better fit to this regression model.

*6.3 Impact of Industry-Specific Factor on Profitability (ROA)*

This study reports that industry-specific factor i.e. market concentration has significant impact on profitability performance of all listed commercial banks of Bangladesh. Market concentration measured by HHI for loans found to have positive association with ROA of all listed commercial banks. A unit increase in HHI indicative of lowering of competition in banking sector results in increment in ROA of commercial banks by 0.0000224 units.

*6.4 Impact of Macro-Economic Factors on Profitability*

The effects of macro-economic factors that are examined in this study are real inflation rate and GDP growth rate. Both indicate negative impact on profitability, however, none of these macro-economic factors have statistically significant impact on listed commercial banks of Bangladesh.

*6.5 Impact of Banking System on Profitability*

The profitability determinants are same for both Conventional and Islamic banks according to afore stated regression analysis. However, the ROA of Conventional and Islamic banks are not same. They are significantly different from each other due to the impact of differing banking system. ROA of Islamic bank is higher than ROA of conventional banks of Bangladesh by 0.0016984 units as indicated in the regression result. Thus, it is established in this study that differences in banking system exerts significant effect on earnings commercial banks of Bangladesh.

## 7. Conclusion and Policy Implications

Evaluation of profitability of listed commercial banks of Bangladesh reveals that on an average, profitability is exhibiting a decreasing trend over the selected period. The profitability performance of Islamic banks was rather high but volatile compared to Conventional banks. This result is consistent with the earlier researches by Rosly and Bakar (2003); Mahmood (2005), Olson and Zoubi (2008) and Usman and Khan (2012). All these researches found Islamic banks to have higher profitability compared to Conventional banks. This study finds that apart from macro-economic variables bank-specific, industry-specific and the different banking system plays a significant role in explaining profitability of the banking sector of Bangladesh. The regression model examining impact of selected determinants on profitability is statistically significant as a whole indicating goodness of fit of the model. Hence, the variables used in the corresponding regression line for ROA are better able to explain variations in this dependent variable than without having any variables in the regression line.

Evaluation of profitability of commercial banks of Bangladesh created scope for the researcher to propose few suggestions for the interest of academic research as well as for the bank supervisors. This study finds that size of bank affects the profitability of commercial banks of Bangladesh positively. In order to realize the benefit of economies of scale deriving from increase in bank size, bank management should pay close attention to controlling coordination; and increased communication and transportation costs while managing large-scale and diverse operations. Credit risk and liquidity risk is found to be an important determinant of profitability of commercial banks of Bangladesh since credit quality is considered as one of the main indicators of financial soundness and health of bank. To make efficient lending decision banks should choose borrowers who have relatively low credit risk or high collateral to back their loans. Moreover, to monitor lending decision banks must rely on collecting both soft and hard information from borrowers and also should focus on relationship development with borrower. This study finds that banking sector of Bangladesh enjoys high level of liquidity in terms of having high volume of loans and advances in its portfolio against its deposits. However, liquidity of loans and advances depends on the recovery of such earning assets; hence credit risk plays an important role though banks apparently maintaining high liquidity. Thus, efficient and informed lending decisions will help bank not only to minimize default risk but also to maintain liquidity of its earning assets. This study finds that higher equity capital helps banks to earn higher profits. This study

International Journal of Accounting and Financial Reporting
ISSN 2162-3082
2018, Vol. 8, No. 4
advocates holding of high level of equity capital by the commercial banks of Bangladesh because it discourages bank management to take excessive risk at the expense of liability holders through high leverage, thus reducing the risk-shifting moral hazard problem of bank managers. Besides, when banks hold more than the regulatory required capital, they can channel this excess capital and invest it in the form of loans and securities and on any portfolio of risky assets, and thereby can earn higher profits. In addition, this study finds that profitability of commercial banks of Bangladesh is affected by the inefficient cost management by the banks. However, as profitability of commercial banks is also shaped by its size, banks can be more profitable by exerting market power through taking advantage of its size and also by having operational costs efficiency. Hence, opportunity exists for cost reduction through scale economies, however, to reap benefit from such opportunities effective cost control mechanism must be in place or bank may employ the best cost-control practices of the industry to avoid inefficient cost management.

Banking sector of Bangladesh is experiencing moderate concentration with fair level of competition prevailing among the banking units. However, the falling HHI is a sign of rising competition in the banking sector. This study finds positive association of HHI with profitability that means large banks are still able to extract monopolistic rents by using their market power to charge higher interest rates on loans and offering lower rates on deposits in the banking sector of Bangladesh. However, this sort of price setting in banking sector is not expected by the customers, hence opportunity prevails for small banks to attract the customers because the gradual decline in HHI is lowering the concentration and increasing the competition in the banking sector; and competition is expected to increase further in future. Hence banks should focus on expanding its operation to exert market power and also to reap benefit of scale economies by employing effective cost-control mechanism. Moreover, with the increased size banks should focus more on coordination and offer differentiated products and services; and also must excel in service delivery to retain the existing customer and attract new customers. This study reports that the differences in banking system exert significant influence on profitability of commercial banks. Hence, it is expected that the policies and directives of Bangladesh Bank would be directed towards enhancing the profitability of both types of banks with the aim of achieving major financial goals of profit maximization in order to build resilience and retain stability in the country's financial system.

## Acknowledgement

This research is part of PhD thesis of Dr. Nusrat Jahan conducted under the supervision of former Professor K.M. Gloam Muhiuddin of Accounting Department, Chittagong University, Chittagong, Bangladesh.
## References

Alkassim, F. A. (2005). The profitability of Islamic and conventional banking in the GCC countries: a comparative study. *Journal of Review of Islamic Economics, 13*, 5-30.

Athanasoglou, *et al.* (2006). Determinants of bank profitability in the South Eastern European region. *Journal of Financial Decision Making, 2,* 1-17.
176                                                                           http://ijafr.macrothink.org

**International Journal of Accounting and Financial Reporting**
ISSN 2162-3082
2018, Vol. 8, No. 4

Van Schaik, D. (2001). Islamic Banking. *The Arab Bank Review, 3*(1), 45-51.

**Glossary**

NCBs = Nationalized Commercial Banks

SCBs = Specialized Commercial Banks

PCBs = Private Commercial Banks

FCBs = Foreign Commercial Banks

DFIs= State-owned Development Financial Institutions

TFP = Total Factor Productivity

Crisk = Credit Risk

Liq = Liquidity Risk

Opeff = Operating Inefficiency

Capstr = Capital Structure

HHI= Herfindahl–Hirschman Index

Infl= Inflation Rate

Bn= Bank Name

Bc= Bank Category